\documentclass[manuscript,screen]{acmart}

\AtBeginDocument{%
	\providecommand\BibTeX{{%
			\normalfont B\kern-0.5em{\scshape i\kern-0.25em b}\kern-0.8em\TeX}}}

\usepackage{fancyhdr}
\usepackage[ruled, linesnumbered]{algorithm2e}
\usepackage{graphicx}
\usepackage{subfigure}
\usepackage{url}
\usepackage{multirow}
\usepackage{array}
\usepackage{booktabs}
\usepackage{enumitem}
\usepackage{rotating}

\usepackage{amsmath,amssymb,amsthm}  

\usepackage{xcolor}

\usepackage{listings}
\usepackage{caption}

\setcopyright{acmcopyright}
\copyrightyear{2022}
\acmYear{2022}
\acmDOI{XXXXXXX.XXXXXXX}

\setcopyright{none}
\renewcommand\footnotetextcopyrightpermission[1]{} 
\acmConference[TOSEM]{}{2022}{}
\acmPrice{15.00}
\acmISBN{978-1-4503-XXXX-X/18/06}

\begin{document}

\title{1-to-1 or 1-to-n? \\ Investigating the effect of function inlining on binary similarity analysis}

\author{Ang Jia}
\email{jiaang@stu.xjtu.edu.cn}
\author{Ming Fan}
\email{mingfan@mail.xjtu.edu.cn}
\author{Wuxia Jin}
\email{jinwuxia@mail.xjtu.edu.cn}
\author{Xi Xu}
\email{xx19960325@stu.xjtu.edu.cn}
\author{Zhaohui Zhou}
\email{zhzhou@stu.xjtu.edu.cn}
\affiliation{%
	\institution{Xi'an Jiaotong University}
	\city{Xi'an}
	\country{China}
}

\author{Qiyi Tang}
\email{(dodgetang@tencent.com}
\author{Sen Nie}
\email{snie@tencent.com}
\author{Shi Wu}
\email{shiwu@tencent.com}
\affiliation{%
	\institution{Tencent Security Keen Lab}
	\city{Shanghai}
	\country{China}
	\postcode{710049}
}

\author{Ting Liu}
\email{tingliu@mail.xjtu.edu.cn}
\affiliation{%
	\institution{Xi'an Jiaotong University}
	\city{Xi'an}
	\country{China}
}

\renewcommand{\shortauthors}{Ang Jia et al.}

\begin{abstract}

Binary similarity analysis is critical to many code-reuse-related issues, where function matching is its fundamental task. ``\textbf{1-to-1}'' mechanism has been applied in most binary similarity analysis works, in which one function in a binary file is matched against one function in a source file or binary file. However, we discover that the function mapping is a more complex problem of ``\textbf{1-to-n}'' (one binary function matches multiple source functions or binary functions) or even ``\textbf{n-to-n}'' (
multiple binary functions match multiple binary functions) due to the existence of \textbf{function inlining}, different from traditional understanding.

In this paper, we investigate the effect of function inlining on binary similarity analysis. 
To support our study, a scalable and lightweight identification method is designed to recover function inlining in binaries. 
88 projects (compiled in 288 versions and resulting in 32,460,156 binary functions) are collected and analyzed to construct 4 inlining-oriented datasets for various similarity analysis tasks, including code search, OSS (Open Source Software) reuse detection, vulnerability detection, and patch presence test.
Then, we further study the extent of function inlining, the performance of existing works under function inlining, and the effectiveness of existing inlining-simulation strategies. Results show that the proportion of function inlining can reach nearly 70\%, while most existing works neglect it and use ``1-to-1'' mechanism. 
The mismatches cause 30\% loss in performance during code search and 40\% loss during vulnerability detection. Moreover, most inlined functions would be ignored during OSS reuse detection and patch presence test, thus leaving these functions risky.
We further analyze 2 inlining-simulation strategies on our dataset. It is shown that they miss nearly 40\% of the inlined functions, and there is still a large space for promotion.
By precisely recovering when function inlining happens, we discover that inlining is usually cumulative when optimization increases. 
Thus, conditional inlining and incremental inlining are suggested to design a low-cost and high-coverage inlining-simulation strategy. \footnote{New Paper}

\end{abstract}



\maketitle

\section{Introduction}\label{introduction}

According to a report published by Gartner \cite{Gartner_SCA}, over 90\% of the development organizations stated that they rely on open source components. Although proper use of open-source code helps reduce the development cost, improper use of open-source code may bring legal and security risks. Developers may unintentionally violate open-source licenses, consequently causing great financial loss to software companies. For example, Cisco and VMware were exposed to serious legal issues because they did not adhere to the licensing terms of the Linux kernel \cite{Cisco, VMware}. Moreover, vulnerabilities are easy to exploit in open-source software since its code is accessible. Software systems that use the vulnerable code and do not patch in time will suffer from security risks \cite{FIBER}. Binary code similarity analysis works are proposed to resolve these code-reuse-related issues.


The fundamental task of binary code similarity analysis is to match the query binary function and target source/binary functions. Here, the query function is the function in a binary absence of its origin, and the target function represents functions that may be related to a license, vulnerability, or patch. 
Given the query binary function, finding its similar binary functions is called binary2binary matching. Binary2binary matching \cite{DBLP:conf/icics/GaoRS08, DBLP:conf/ndss/ZuoLYL0Z19, DBLP:conf/pldi/DavidPY16, DBLP:journals/it/PewnyGGRH17, DBLP:journals/tse/LuoMWLZ17} has various applications including binary code search \cite{asm2vec, Kam1n0, safe, Gemini}, vulnerability detection \cite{Tracy, vulseeker} and patch presence test \cite{Binxray}.
Given the query binary function, finding its similar source functions is called binary2source matching. Binary2source matching \cite{BAT, RESource, JISIS14, BinPro, OSSPolice, Saner2019, B2SFinder, CodeCMR, buggraph, ban2021b2smatcher} is critical to many code-reuse-related tasks, including binary2source code search \cite{CodeCMR} and OSS reuse detection \cite{B2SFinder, ban2021b2smatcher}. The matching mechanism of both binary2binary matching and binary2source matching is always considered as ``\textbf{1-to-1}'' in most existing works, i.e., one function in a binary file is matched against one function in a source file or binary file. However, we discover that such mapping is usually a more complex problem of ``\textbf{1-to-n}'' or ``\textbf{n-to-n}'' due to the existence of \textbf{function inlining}~\cite{Function_inlining}, which differs from traditional assumptions.


\begin{figure*}[htbp]
	\centering
	\vspace{-5pt}
	\includegraphics[width=0.85\textwidth]{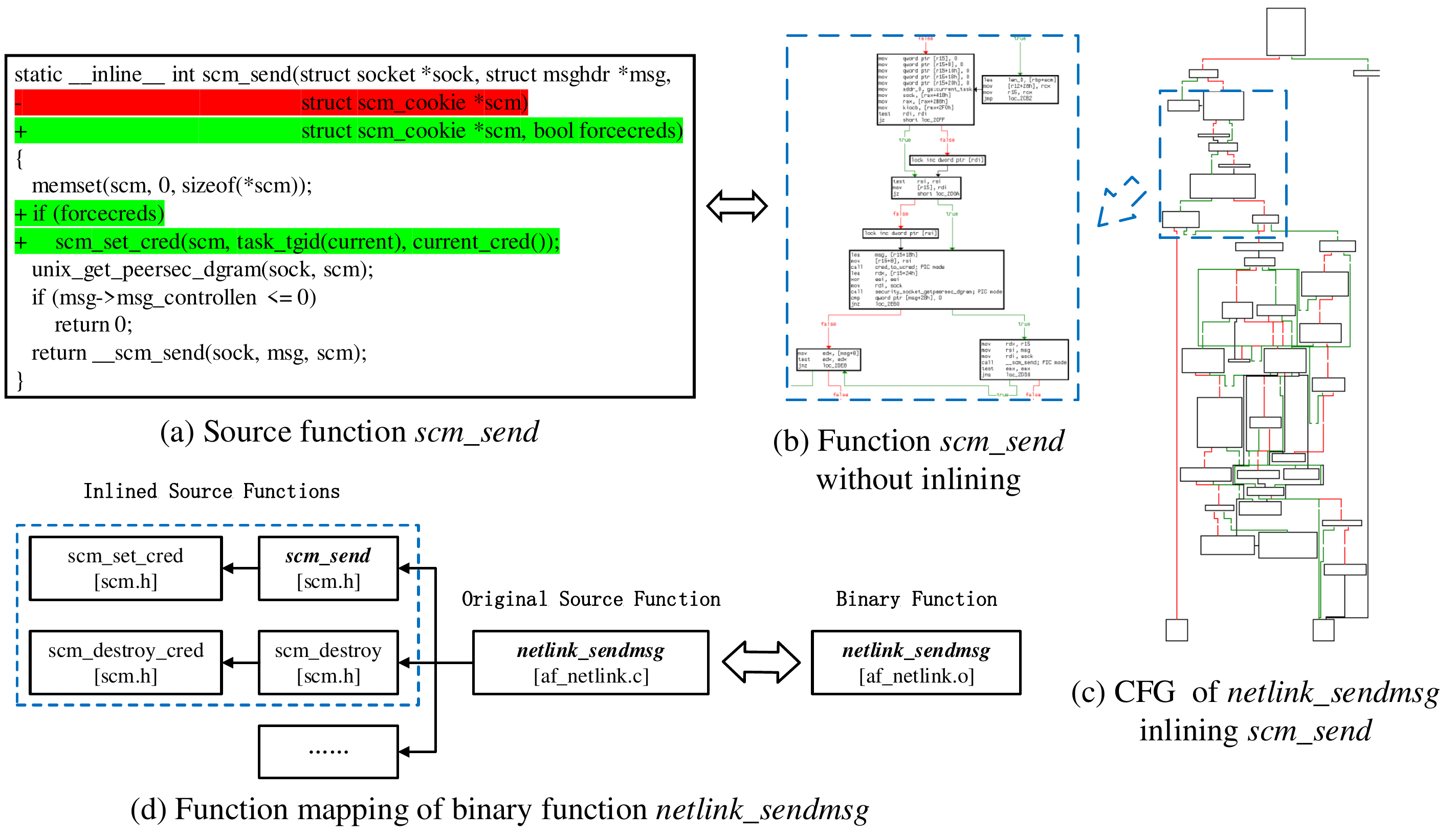}
	\vspace{-8pt}
	\caption{``1-to-n'' matching cases brought by function inlining}
	\label{fig:linux_inline_example_all}
	\vspace{-10pt}
\end{figure*}

Figure \ref{fig:linux_inline_example_all} shows a motivating example when detecting vulnerable functions under inlining.
Figure \ref{fig:linux_inline_example_all} (a) shows the source code of function \textit{scm\_send} in vulnerable version and patched version in the project Linux \cite{Linux_git} for CVE-2020-16593 \cite{linux_example}. 
As code reuses of this vulnerable function may exist, the vulnerability would be propagated to disparate binaries.
Detecting the influenced binaries requires searching whether there is a function in the binary similar to the vulnerable function in the source code. To support this process, Figure \ref{fig:linux_inline_example_all} (c) depicts the complete CFG (Control Flow Graph) of a binary function \textit{netlink\_sendmsg} that will be checked, with a partial CFG shown in Figure \ref{fig:linux_inline_example_all} (b). Each node represents a basic block, and the array represents the control flow between them. 


Following the ``\textbf{1-to-1}'' mechanism, we cannot find an individual binary function that is matching to the function \textit{scm\_send}. Analyzing the debug information, we found the  \textit{scm\_send}'s code is compiled into another function named \textit{netlink\_sendmsg}. However, the existing works calculate a low similarity score between the function \textit{scm\_send} and the target binary function \textit{netlink\_sendmsg}, indicating they are unmatched and thus leading to a failure that the vulnerability in \textit{netlink\_sendmsg} is missed.


\textbf{Function inlining} causes the failure of the ``1-to-1'' matching mechanism.
Figure \ref{fig:linux_inline_example_all} (d) shows inlining happen to function \textit{netlink\_sendmsg}. Function \textit{scm\_send}, called by \textit{netlink\_sendmsg}, is inlined into \textit{netlink\_sendmsg}.
When a function inlining occurs, it replaces the call-site statement inside a caller with the function body of the callee. As a result, the produced binary function will contain multiple source functions. Even more, the function inlining can be \textbf{nested}, meaning that an inlined function may further inline another function. As shown in Figure \ref{fig:linux_inline_example_all} (d),  the \textit{scm\_send}, inlined by \textit{netlink\_sendmsg}, inlines \textit{scm\_set\_cred}.  We can observe that the mapping between binary function \textit{netlink\_sendmsg} and its original source functions is no longer a ``1-to-1'' mapping but a ``1-to-n" issue --- \textit{one binary function maps multiple source functions, sometimes even in a nested inlining manner}. Thus, the existing ``1-to-1'' matching mechanism, which tries to match two functions with unequal semantics, suffered the expected failure.

``1-to-n" issues bring a great challenge to existing binary2source matching works, as we have shown. Caused by the function inlining, the function mapping in binary2binary matching has become ``1-to-n'' and even ``n-to-n''. However, to the best of our knowledge, few existing works have systematically studied the effect of function inlining on binary2source and binary2binary matching tasks (together, we call binary code similarity analysis). It is still unclear the extent of function inlining, and its impact on existing binary similarity analysis works. This paper will address this issue, and concretely we will investigate the three research questions.


\emph{\textbf{RQ1}: To what extent will inlining happen during compilation?}
To answer this RQ, we will evaluate the frequency and degree of function inlining, which can demonstrate the wild presence of function inlining in binary code similarity analysis tasks. This RQ will be answered in Section \ref{sec:inline_statistics}.

\emph{\textbf{RQ2}: What is the effect of function inlining on the performance of existing binary code similarity analysis works?}
To answer this RQ, we will evaluate the performance of existing works that do not consider function inlining or do not fully handle function inlining (which account for the majority). The decrease in performance indicates the importance of considering function inlining. We will answer this RQ in Section \ref{sec:experiments}.

\emph{\textbf{RQ3}: Can existing inlining-simulation strategies solve them? 
}
Existing works have proposed strategies to simulate the results of function inlining (here, we call them inlining-simulation strategies). To answer this RQ, we will evaluate the effectiveness of existing inlining-simulation strategies from the inlining cost and coverage of inlined functions. The shortcomings of existing inlining-simulation strategies indicate directions for improvement. This RQ will be answered in Section \ref{sec:inline_pattern}.

We first construct four inlining-oriented datasets for four binary code analysis tasks, including code search, OSS reuse detection, vulnerability detection, and patch presence test to support our study. Furthermore, we propose a scalable and lightweight identification method that can automatically recover function inlining in binaries. 
Based on our labeled dataset, we then investigate the extent of function inlining, the impact of function inlining on existing binary code similarity works, and the effectiveness of existing inlining-simulation strategies. Finally, we discuss the interesting findings and give our suggestions for designing more effective inlining-simulation strategies.

Our main contributions are listed below:
\vspace{-5pt}
\begin{itemize}
	\item To the best of our knowledge, we are the first to comprehensively evaluate the effect of function inlining on binary code similarity analysis tasks.
	
	\item We create the first function inlining dataset by designing a scalable and lightweight identification method that can automatically recover function inlining in binaries (Section \ref{sec:dataset}).
	
	\item Our study recovers the high frequency and high degree  of function inlining (Section \ref{sec:inline_statistics}). Averagely the proportion of function inlining can range from 40\% to 70\% and a binary function can inline 2-4 source functions.
	
	\item We conduct five experiments to evaluate the performance of existing binary code similarity works under function inlining (Section \ref{sec:experiments}). Code search suffers a 30\% decline in detecting functions with inlining while OSS reuse detection and patch presence test miss most inlined functions.
	
	\item And we take a deep analysis of existing inlining-simulation strategies (Section \ref{sec:inline_pattern}) and point out the directions for designing more effective strategies (Section \ref{sec:suggestion}). 
	
\end{itemize}

To facilitate further research, we have made the source code and dataset publicly available \cite{OpenSource}.

\vspace{-5pt}
\section{Preliminary}
\label{sec:preliminary}

This section will introduce existing static binary code similarity analysis works, commonly-used inlining strategies, and the research scope of our work.

\vspace{-5pt}
\subsection{Binary Code Similarity Analysis}

\subsubsection{\textbf{Binary2binary code matching}}

According to the granularity to conduct matching, existing binary2binary matching can be classified into instruction level, basic block level, and function level method.

Bindiff\cite{bindiff}, Kam1n0\cite{Kam1n0} and Asm2Vec\cite{asm2vec} are instruction level methods. To conduct matching, Bindiff and Kam1no are working on the raw assembly instructions and conducting sub-graph matching to find similar functions. Instead, Asm2Vec trains the embedding for the assembly code and matching functions using function-level embedding by lifting instruction-level embedding.

Deepbindiff\cite{deepbindiff} is a basic block level method to conduct binary diffing. To conduct matching, Deepbindiff first constructs a connected control flow graph of two binaries by merging their common-used library calls and strings. Then it applies TADW\cite{TADW} to learn embeddings of basic blocks. Finally, they obtain the matching by calculating the similarities of embeddings.

Function level matching methods \cite{bingo, bingo-E, Tracy, Gemini, safe, vulseeker, Binxray} are the most widely-used methods when conducting binary2binary code matching. Tracy \cite{Tracy} extract the data flow strands from functions and matching functions by the hash of strands. Gemini\cite{Gemini} uses the CFG (control flow graph) augmented with block attributes as the representation of a function and trains a Siamese network for function matching. SAFE \cite{safe} use a self-attentive network \cite{Self-attentive_network} directly on instruction embeddings to obtain the function embeddings. Vulseeker \cite{vulseeker} works on labeled semantic flow graphs (LSFG) and uses its embedding to find vulnerable functions. BinXray \cite{Binxray} uses diffs between vulnerable version and patch version to conduct path presence test. 

Among these binary2binary matching works, Bingo \cite{bingo, bingo-E} and Asm2Vec \cite{asm2vec} especially propose selective inlining-simulation strategies to tackle the challenges that function inlining brings. However, our work is different from theirs. We aim to provide a systematical and general understanding of function inlining and its impacts. Moreover, in our work, we also evaluate the effectiveness of their strategies leveraging our unique insights of function inlining.

\subsubsection{\textbf{Binary2source code matching}}

Existing binary2source code matching works can be classified into feature-based methods and learning-based methods. 

Feature-based methods \cite{BAT, RESource, JISIS14, BinPro, OSSPolice, Saner2019, B2SFinder} are commonly used by early works. BAT \cite{BAT} considers textual strings as features to match source and binary code. BinPro \cite{BinPro} extracts similar features but utilizes machine learning algorithms to compute optimal code features for the matching. B2SFinder \cite{B2SFinder}, a state-of-the-art feature-based method, considers three types of features, including string, integer, and control-flow features. As these features are sometimes rare at the function level, most feature-based methods conduct matching at the file level.

CodeCMR \cite{CodeCMR} is a learning-based method for function-level binary2source matching. CodeCMR models binary2source matching as a cross-modal retrieval task, with the source function code and binary function code as two neural network inputs. The model output is the similarity between them. CodeCMR also extracts string and integer features to supply the inputs. It has achieved fast and precise matching by developing a deep learning method. 

\vspace{-5pt}
\subsection{Function Inlining}
\label{function_inline}

Function inlining \cite{Function_inline} conducts function expansion, replacing a function call with the callee body. Besides that function inlining benefits decreasing the cost of method calls, its major purpose is to enable additional compilation optimizations  \cite{DBLP:conf/cgo/ProkopecDLW19}. Though there are passes that can be used for inter-procedural optimizations, inlining the callee function can convert inter-procedural optimizations into intra-procedural optimizations, facilitating more optimizations.

On the other hand, function inlining can cause an increase in binary code size and compilation time. It is a trade-off to determine when to inline by a balance between these benefits and costs. Prior researches have proposed lots of inlining strategies \cite{DBLP:journals/tse/DavidsonH92, DBLP:conf/lfp/DeanC94, hubicka2004gcc, DBLP:conf/lcpc/ZhaoA03, DBLP:conf/cc/CooperHW08, andersson2009evaluation, DBLP:conf/pldi/HwuC89, durand2018partial} from different aspects. We will introduce inlining strategies adopted in GCC \cite{gcc} and LLVM \cite{LLVM}, two of the most popular compilers.

GCC and LLVM follow a similar workflow for function inlining, including inlining decision and inlining conduction. To decide which call-sites should be inlined, they consider user-forced inlining, user-suggested inlining, and normal inlining candidates. User-forced inlining, such as ``$\_\_attribute\_\_(always\_inline)$'', will first be considered to be inlined, as long as the call-site of this function does not violate fundamental requirements. User-suggested inlining is specified by users using the keyword ``inline'' to decorate function definitions. Such decorated functions will be possible candidates to be inlined, but whether to inline still depends on the strategy of the compiler. Normal inlining candidates will be identified and determined according to the strategies used in GCC and LLVM themselves.

GCC uses a priority-guided inlining algorithm to decide which call-sites should be inlined. They first define the inlining priority of a function call by pre-estimating the benefit of this inlining. Then they conduct inlining following the priorities. At the same time, the cost of inlining also increases. Once the total cost of inlined methods reaches a maximal threshold as configured, the inlining process will stop.

Instead, LLVM uses a prediction-driven inlining method to determine function inlining. It first classifies functions into hot and cold ones according to the times they are called. Then each call-site will be assigned with measurements of inlining benefit and inlining cost \cite{InlineCost.cpp}. If the cost value is lower than the benefit value, this call-site will be inlined \cite{LLVM_inline}.

Besides, compilers also implement specific settings for small-size functions and functions that are called only once. For example, when compiling a program with the option of ``-O1", GCC-10 applies ``-finline-functions-called-once'' to conduct inlining. This option will inline all static functions called only once no matter whether they are marked with ``inline'' or not. Moreover, if a function called once is inlined, then the function will not be compiled as an individual function in the binary\cite{GCC_inline}, such as the function \textit{scm\_send} in Figure \ref{fig:linux_inline_example_all}.

\subsection{Research Scope}
Before introducing our work, we first clarify our research scope. Our work investigates the impact of inlining strategies used in common compilers on static function-level binary code similarity analysis works. Other function inlining strategies or instruction level, basic block level and file level method is beyond our scope.

We focus on C/C++ programs compiled by GCC and Clang with O0-O3. The main architecture of binaries is in X86-64, but we also analyze datasets compiled in 8 architectures. Moreover, we only consider the inlining of user-defined functions in this paper. The inlining of library functions is similar to user-defined functions, but library functions may differ when the compilation environment changes.


\section{Dataset construction}
\label{sec:dataset}

\begin{figure*}[htbp]
	\centering
	\includegraphics[width=0.65\textwidth]{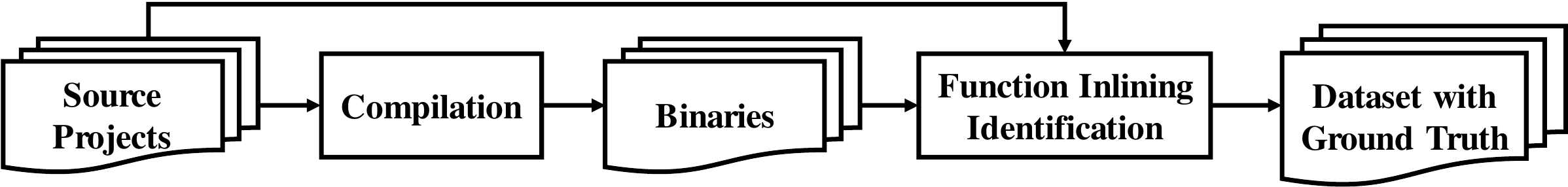}
	\vspace{-8pt}
	\caption{Workflow of dataset construction}
	\label{fig:dataset_construction}
	\vspace{-12pt}
\end{figure*}

As few studies have researched the impact of function inlining on binary code similarity analysis work, an inlining-oriented dataset is needed. To fill the gap, we will first collect 88 projects for different applications and propose a scalable and lightweight identification method to recover function inlining in binaries automatically. As illustrated in Figure \ref{fig:dataset_construction}, we first collect source projects and compile them into binaries under diverse configurations (Section \ref{sec:dataset_construction}). Based on the source code and binaries, we propose an automatic identification method to identify function inlining in binaries (Section \ref{sec:dataset_labeling}).  

\subsection{Dataset Collection}
\label{sec:dataset_construction}

In this work, we focus on C/C++ programs and two popular compilers, GCC \cite{gcc} and Clang \cite{clang}. We have collected four datasets by employing and extending Binkit \cite{Binkit, Binkit_git}, a binary code similarity analysis benchmark. The compositions and compilation settings of these dataset are listed in Table \ref{tab:composition_of_four_dataset}.

\begin{table}[h]
    \vspace{-5pt}
    \caption{Composition of four datasets}
    \vspace{-8pt}
    \begin{tabular}{c|c|c|c|c|c}
        \hline
        Dataset & \# of Projects & Representative Projects & Compilation Setting                                                                                 & Versions & Binary Functions \\ \hline
        I       & 51             & Coreutils\cite{coreutils}, Findutils\cite{findutils}    & Table \ref{tab:datasetc_compiler_setting}                                                                                            & 288      & 18,783,986       \\ \hline
        II      & 24             & bzip2\cite{bzip2}, minizip\cite{minizip}         & gcc-8.2.0, X86-64, O0-O3                                                                            & 4        & 190,611          \\ \hline
        III     & 12             & Openssl\cite{openssl}, Binutils\cite{binutils}       & \begin{tabular}[c]{@{}c@{}}gcc-8.2.0, X86-64, O0-O3,\\ inline, noinline\end{tabular} & 8        & 8,635,362        \\ \hline
        IV      & 1              & Chromium\cite{chromium}              & clang, debug, release                                                                               & 2        & 4,850,197       \\  \hline
    \end{tabular}
    \label{tab:composition_of_four_dataset}
    \vspace{-5pt}
\end{table}

\begin{table}[htbp]
	\vspace{-5pt}
	\caption{Compilers, optimizations and architectures used in dataset I}
	\label{tab:datasetc_compiler_setting}
	\vspace{-8pt}
	\centering
	\begin{tabular}{c|c}
		\hline
		Compilers     & \begin{tabular}[c]{@{}c@{}}gcc-4.9.4, gcc-5.5.0, gcc-6.4.0,  gcc-7.3.0, gcc-8.2.0, clang-4.0, clang-5.0, clang-6.0, clang-7.0\end{tabular}         \\ \hline
		Optimizations & O0, O1, O2, O3                                                                                                                                        \\ \hline
		Architectures & \begin{tabular}[c]{@{}c@{}}X86-32, X86-64, ARM-32, ARM-64,  MIPS-32, MIPS-64, MIPSeb-32, MIPSeb-64\end{tabular} \\ \hline                     
	\end{tabular}
	\vspace{-5pt}
\end{table}

\textbf{Dataset I - for code search.} Dataset I was comprised by the 51 packages from GNU projects\cite{GNU_project}, compiled by 9 compilers, 4 optimizations, to 8 architectures (shown in Table \ref{tab:datasetc_compiler_setting}), resulting in 288 compilation versions containing 67,680 binaries and 18,783,986 functions. Dataset I is collected from Binkit\cite{Binkit}. This dataset contains many widely-used packages such as \textit{Coreutils} \cite{coreutils} which have been extensively used in binary similarity detection works \cite{xu2021interpretation, bingo, liu2018alphadiff, egele2014blanket, wang2017memory, CodeCMR}.

\textbf{Dataset II - for OSS reuse detection.} Dataset II was extended from a partial reuse dataset used in ISRD\cite{xu2021interpretation} containing 24 programs with 74 partial reuses. It was originally compiled by GCC with O2 in X86-64. We extend it to 4 optimizations by using gcc-8.2.0. 
Dataset II is not extended to other architectures because existing OSS reuse detection works do not include them.
Dataset II was comprised of 96,085 source functions and 190,611 binary functions.

\textbf{Dataset III - for vulnerability/patch detection.} Dataset III was constructed from the dataset used in BinXray\cite{Binxray}, containing 479 CVEs and 6238 vulnerable functions. We generate two versions of Dataset III, including Dataset III (without additional compilation flags) and Dataset III-NI (with flag ``\textit{-fnoinline}'' to forbid inlining). Dataset III is comprised of 8,635,362 binary functions.

\textbf{Dataset IV - for application binaries.} Apart from these three datasets, we also construct a dataset by compiling chromium\cite{chromium} with Clang in its debug and release optimization to X86-64 to form the dataset IV. Dataset IV is comprised of 4,850,197 binary functions. This dataset will provide a vision of function inlining in large application binaries.

\subsection{Function Inlining Identification}
\label{sec:dataset_labeling}
In this section, we will first illustrate terminologies that we will use throughout our paper and then introduce our function inlining identification method.

\subsubsection{Terminology}

To make the representation more clear, we first illustrate our notions of the functions from both source and binary. Usually, as shown in Figure \ref{fig:linux_inline_example_all} (d), when function inlining happens, a binary function (\textit{netlink\_sendmsg}) is compiled from the original source function (\textit{netlink\_sendmsg}) and some inlined source functions (\textit{scm\_send}, \textit{scm\_set\_cred} and etc.). Here, we named the produced binary function (\textit{netlink\_sendmsg}) as BFI (binary function with inlining), the original source function (\textit{netlink\_sendmsg}) as OSF, and the inlined source functions (\textit{scm\_send}, \textit{scm\_set\_cred} and etc.) as ISFs. Similarly, binary function without inlining is named as NBF (normal binary function), NBF mapped source function as NSF (normal source function). Moreover, binary function as BF and source function as SF. These notions will be used throughout the paper.

\subsubsection{Function inlining identification method.} To identify function inlining, we first construct function mappings from source code to binaries, which can also serve as the ground truth for binary code similarity analysis.
When compiling a program in debug mode, compiler will record the mapping from binary instructions to source instructions. Following that mapping, we further find the binary function that binary instructions belong to and the source function that source instructions belong to. Thus, function-to-function mapping can be generated.

\begin{figure}[t]
	\centering
	\vspace{-5pt}
	\includegraphics[width=0.7\textwidth]{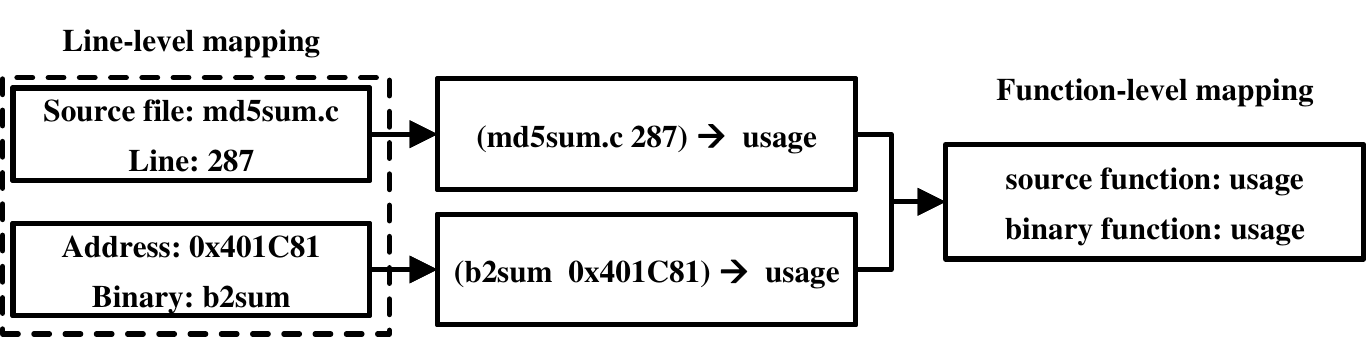}
	\vspace{-8pt}
	\caption{Example of extending line-level mapping to function-level mapping}
	\label{fig:dataset_labeling_example}
	\vspace{-12pt}
\end{figure}

Algorithm \ref{alg:dataset_labeling} formalizes our method for generating function-level mappings for datasets. The method leverages the line mapping implied by the relation between binary address and source code line, provided in the debug mode. By extending line mapping results, the method will generate a function-level mapping between binary functions and source functions. We will introduce the algorithm with an example in Figure \ref{fig:dataset_labeling_example}.

\begin{algorithm}[t]
    \caption{Method to generate function-level mappings}
    \label{alg:dataset_labeling}
    \small
    \KwIn{Source project $S$, Binary with debug info $B$}
    \KwOut{Function mapping $FM_{b2s}$}
    \SetKwProg{Fn}{Function}{:}{} 
    \SetKwFunction{FMain}{Main}
    \SetKwFunction{func}{Preprocessing}
    \Fn{\func{$B$, $S$}}{
        $LM_{b2s}$ = Extract\_Address\_to\_Line\_Mapping($B$)     \tcp*[h]{return a list of (address, line) mapping}  \label{alg:b2s}
        
        $SM_{l2f}$ = Extract\_Line\_to\_SF\_Mapping($S$) \tcp*[h]{return a dict of line to function mapping}  \label{alg:l2f}
        
        $BM_{a2f}$ = Extract\_Address\_to\_BF\_mapping($B$)
        \tcp*[h]{return a dict of address to function mapping}\label{alg:a2f}
        
        \KwRet $LM_{b2s}$, $SM_{l2f}$, $BM_{a2f}$\;
      }
    $LM_{b2s}$, $SM_{l2f}$, $BM_{a2f}$ = \func{$B$, $S$}\;
    $FM_{b2s} = \{\}$\;
    \For(\tcp*[h]{traverse all address-to-line mappings}){(address, line) in $LM_{b2s}$ \label{alg:start}}   
    {
        $SF = SM_{l2f}[line]$\;
        $BF = BM_{a2f}[address]$\;
        \If{$BF$ not in $FM_{b2s}$}
        {
            $FM_{b2s}[BF] = []$\;
        }
        \If{$SF$ not in $FM_{b2s}[BF]$}
        {
            $FM_{b2s}[BF].append(SF)$ \tcp*[h]{add mapped source functions to the binary function dict}
        }
    }\label{alg:end}
    \KwRet $FM_{b2s}$ \;
\end{algorithm}

Our method first compiles source code with the ``-g'' option and leverages Dwarf \cite{Dwarf_v5} to produce the .debug\_line section in binary. To extract the line-level mapping, we uses Readelf \cite{readelf} to parse binaries (line \ref{alg:b2s}). As shown in Figure \ref{fig:dataset_labeling_example}, the line mapping contains the mapping between source file lines and binary addresses. 

Next, the method extracts line-to-function mapping in the source code by employing Understand \cite{understand} and extracts address-to-function mapping in the binary code by employing IDA Pro \cite{IDAPro}. Understand and IDA Pro are two state-of-art commercial tools and have been widely used in academia~\cite{pinna2019massive, deepbindiff, pang2021sok} and industry~\cite{understand_case, IDA_case1, IDA_case2}. Understand is a static analysis tool to identify source entities (e.g., files, functions) and dependencies. By exploring source entities, we can infer which entity the code line belongs to (line \ref{alg:l2f}). IDA Pro recovers the function boundary \cite{disassembly_tools} from binaries, outputting binary address to binary function (line \ref{alg:a2f}). For example in Figure \ref{fig:dataset_labeling_example}, source line 287 belongs to the source function \textit{usage}, and binary address 0x401C81 belongs to binary function \textit{usage}. 

Finally, our method extends the above line-level mapping results to function-level mapping by combining these three mappings (lines \ref{alg:start}-\ref{alg:end}). As shown in Figure \ref{fig:dataset_labeling_example}, now we can get the function mapping between source function \textit{usage} and binary function \textit{usage}. 

Our datasets collected in Section 3.1 are labeled by our automatic method. According to the mapping results between BFs and SFs, we can easily identify BFs produced by function inlining (i.e., BFI). Simply, we identify BFs mapped to more than one SFs as BFIs and those mapped to one SF as NBFs. By identifying the OSF or NSF from which BFs are compiled, we can also easily establish binary2binary function-level mappings. Our datasets with function-level mappings can support the study in our work. 

Besides, our method does not rely on any modification of the compilers and can be applied to any binaries compiled in debug mode. Note that Understand and IDA Pro can be substituted with other tools with similar functionalities. We also implement a tool using tree-sitter\cite{tree-sitter} and Ghidra\cite{Ghidra}, facilitating free usages.


\vspace{-5pt}
\section{Investigation on the extent of Function inlining}
\label{sec:inline_statistics}

Before analyzing the extent of inlining, we first discover that inlining is not the same in all circumstances. Firstly, inlining is defined by the compiler. Thus, inlining is influenced by the compilation settings. Secondly, the compiler decides whether to inline by considering the features of a call site, which will differ in different projects. Thus inlining also depends on the projects. Finally, we noticed that different applications also care about different parts of inlining. Thus, in this section, we propose three research questions (RQs) to help understand the extent of function inlining. 








\subsection{\textbf{RQ1.1:} What Is the Extent of Function inlining under different compilations?}

\begin{figure*}[t]
	\centering
	\vspace{-10pt}
	\subfigure[Clang]{
		\begin{minipage}[t]{1\linewidth}
			\centering
			\includegraphics[width=0.65\textwidth]{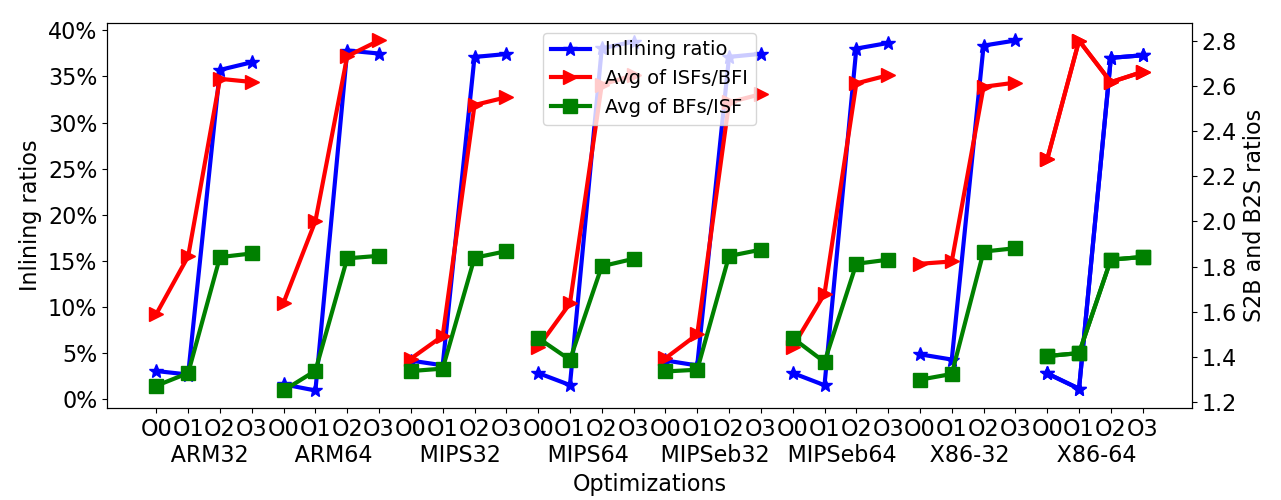}
			\label{fig:compilation_for_clang}
			\vspace{-18pt}
		\end{minipage}%
	}%
	
	\centering
	\vspace{-10pt}
	\subfigure[GCC]{
		\begin{minipage}[t]{1\linewidth}
			\centering
			\includegraphics[width=0.65\textwidth]{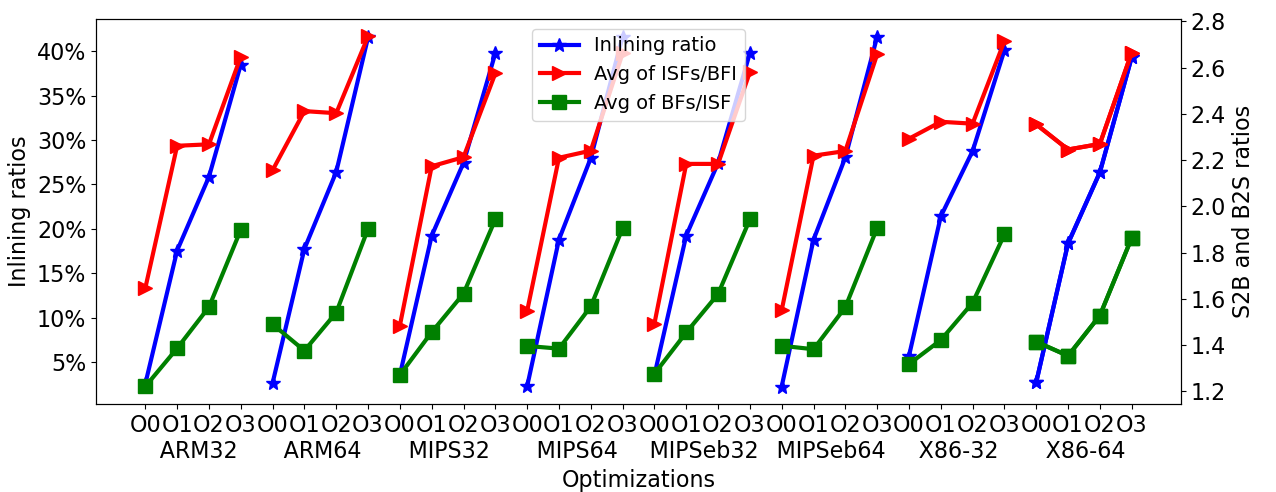}
			\label{fig:compilation_for_gcc}
			\vspace{-18pt}
		\end{minipage}%
	}%
	\vspace{-15pt}
	\caption{Function inlining statistics in different compilation settings}
	\label{fig:dataset_I_statistics}
	\vspace{-17pt}
\end{figure*}

Figure \ref{fig:dataset_I_statistics} shows the statistics of function inlining in dataset I compiled by 8 architectures, 9 compilers, and 4 optimizations. Due to page limit, we only present results in clang-7.0 and gcc-8.2.0. In Figure \ref{fig:dataset_I_statistics}, ``Inlining ratio'' shows the percent of BFIs in BFs, ``ISFs/BFI'' represents the number of ISFs that a BFI inlines, ``BFIs/ISF'' represents the number of BFIs that an ISF is inlined into. Here we calculate these values by average. These metrics measure three aspects associated with function inlining: happening frequency in binary functions (Inlining Ratio), inlining degree in binary functions (ISFs/BFI), and inlining degree in source functions (BFIs/ISF).

\subsubsection{\textbf{Statistics in default Ox optimizations}} For statistics of all 64 compilation options, we can see that the inline ratios range from 0.94\% compiling with clang-7.0 in O1 to ARM64 to 41.67\% compiling with gcc-8.2.0 in O3 to MIPS64. And when a binary function is generated with function inlining, at least 1.39 ISFs are inlined averagely. Moreover, according to our observation, a BFI may inline 200 different source functions to form the final binary function. When we turn the view to BFIs/ISF, an ISF will be inlined by at least 1.22 BFIs on average, and in some cases, an ISF may be inlined into 244 BFIs.

\textbf{Statistics of different optimizations.} For compilation settings from the three levels, we start with the most fundamental setting --- optimizations. No matter what arch or compiler is, there is a straight increase in inlining ratios from low optimizations (i.e., O0, O1) to high optimizations (i.e., O2, O3). Same with the inlining ratios, a BFI tends to inline more ISFs, and an ISF tends to be inlined by more BFIs with higher optimizations. When applying the O3 option, more than 36\% of BFs will inline ISFs. As released binaries are often compiled under high optimizations to ensure their efficiency, we believe that function inlining is a common phenomenon. 

\textbf{Statistics of different compilers.} The influence of different compilers is not directly pushed on generated binaries but through its inlining strategy to effect optimizations. An evident difference between GCC and Clang is that when compiling in O1, GCC produced a binary function with higher inlining ratios than Clang. This is because GCC applies "-finline-functions-called-once'' when using O1, as we have illustrated in section \ref{function_inline}. Besides, the inlining ratio of GCC grows gradually from O0 to O3, while ratios of Clang are divided into two parts: low optimizations and high optimizations. This is also caused by the inlining strategies of compilers. GCC gradually adds options to generate O0 to O3 while Clang uses "\textit{always inliner}" pass in O0 and O1, and "\textit{inliner}" pass in O2 and O3.

\textbf{Statistics of different architectures.} When applying high optimizations, all 64-bit architectures have a little higher inlining frequency and degrees than their corresponding 32-bit architectures except X86. Among these 8 architectures, MIPS-64 has the highest inlining ratio, while ARM-64 has the highest inlining degree.

\subsubsection{\textbf{Statistics of using LTO}} 
\begin{figure}[h]
	\centering
	\begin{minipage}[t]{0.33\textwidth}
		\centering
		\vspace{-6pt}
		\includegraphics[width=1\textwidth]{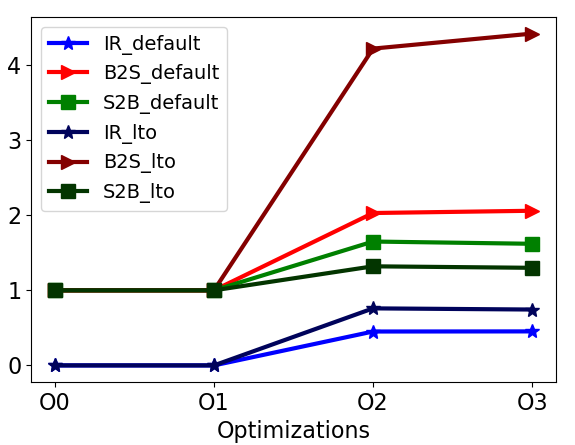}
		\vspace{-10pt}
		\caption{Function inlining statistics with LTO}
		\label{fig:lto}
		\vspace{-7pt}
	\end{minipage}
	\begin{minipage}[t]{0.64\textwidth}
		\centering
		\vspace{-10pt}
		\subfigure[Dataset II]{
			\begin{minipage}[t]{0.48\textwidth}
				\centering
				\includegraphics[width=1\textwidth]{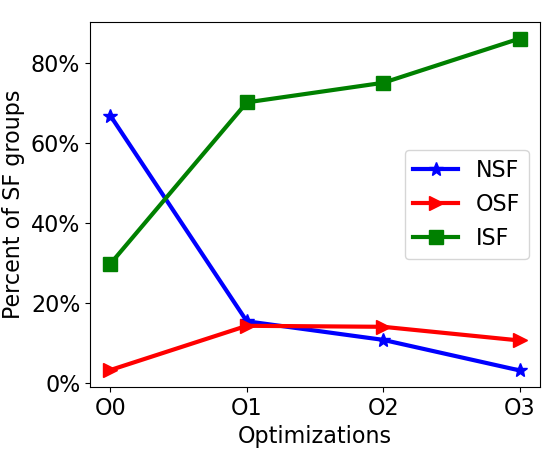}
				\label{fig:inlining_in_II}
				\vspace{-15pt}
			\end{minipage}%
		}%
		\centering
		\vspace{-10pt}
		\subfigure[Dataset III]{
			\begin{minipage}[t]{0.48\textwidth}
				\centering
				\includegraphics[width=1\textwidth]{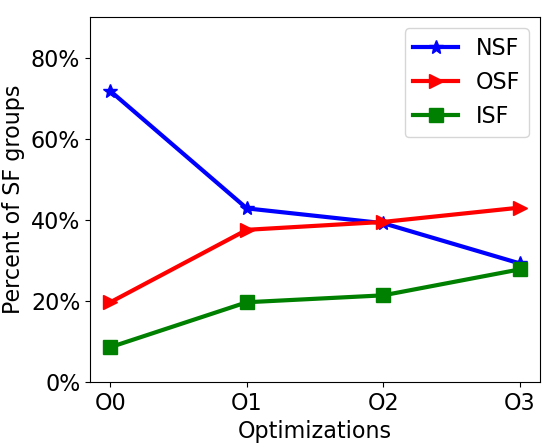}
				\label{fig:inlining_in_III}
				\vspace{-15pt}
			\end{minipage}%
		}%
		\vspace{-5pt}
		\caption{Distribution of critical SF in BFs}
		\vspace{-15pt}
	\end{minipage}
\end{figure}

Apart from the above compilation settings, we also noticed that LTO (Link-Time Optimizations) would also enable inlining across compilation units. Figure \ref{fig:lto} shows the inlining statistics in Coreutils v8.29 using Clang with and without LTO. We use abbreviations to represent the same metrics in Figure \ref{fig:dataset_I_statistics}, such as IR for Inlining Ratio, B2S for ISFs/BFI, and S2B for BFIs/ISF. Compared with binaries without LTO, binaries enabled with LTO have 30\% more inlining ratios. And when compiling in O3 enabled with LTO, more than 70\% of the BFs are BFIs. Apart from the increasing frequency of inlining, 2 more ISFs are inlined into the BFIs when using LTO averagely. The increasing frequency and extent of inlining bring more challenges to existing binary code similarity works.

\textbf{Answering RQ1.1:} 
Inlining happens in high optimizations, and its ratio usually ranges from 30\% to 40\% when applying O3. However, inlining under different compilers may differ at the same optimizations. Besides, when combining O3 with LTO, inlining ratios can raise to 70\%, and averagely a BFI can inline 4 ISFs.



\subsection{\textbf{RQ1.2:} What Is the Extent of Function Inlining in Different Projects?} 

Different projects may have different inlining statistics due to their coding style and design rule. To avoid getting a biased understanding, we first analyze inlining statistics in four kinds of projects, including system code from dataset I, zip/unzip, timing, and data transformation code from dataset II, cryptography, database, and image processing code from dataset III, and browsers from dataset IV.

\begin{table}[h]
	\caption{Function inlining statistics in different kinds of projects}
	\centering
	\vspace{-8pt}
	\scalebox{0.9}{
		\begin{tabular}{c|cccc|cccc|cccc|cc}
			\hline
			D      & \multicolumn{4}{c|}{I}                                                                          & \multicolumn{4}{c|}{II}                                                                         & \multicolumn{4}{c|}{III}                                                                         & \multicolumn{2}{c}{IV}             \\ \hline
			Opt    & \multicolumn{1}{c|}{O0}    & \multicolumn{1}{c|}{O1}     & \multicolumn{1}{c|}{O2}     & O3     & \multicolumn{1}{c|}{O0}    & \multicolumn{1}{c|}{O1}     & \multicolumn{1}{c|}{O2}     & O3     & \multicolumn{1}{c|}{O0}     & \multicolumn{1}{c|}{O1}     & \multicolumn{1}{c|}{O2}     & O3     & \multicolumn{1}{c|}{Og}    & Or     \\ \hline
			IR     & \multicolumn{1}{c|}{2.7\%} & \multicolumn{1}{c|}{18.4\%} & \multicolumn{1}{c|}{26.4\%} & 39.3\% & \multicolumn{1}{c|}{6.5\%} & \multicolumn{1}{c|}{35.6\%} & \multicolumn{1}{c|}{40.2\%} & 48.1\% & \multicolumn{1}{c|}{22.2\%} & \multicolumn{1}{c|}{34.8\%} & \multicolumn{1}{c|}{39.0\%} & 46.5\% & \multicolumn{1}{c|}{6.4\%} & 57.9\% \\ \hline
			B2S\_avg & \multicolumn{1}{c|}{2.35}  & \multicolumn{1}{c|}{2.25}   & \multicolumn{1}{c|}{2.27}   & 2.67   & \multicolumn{1}{c|}{1.86}  & \multicolumn{1}{c|}{2.08}   & \multicolumn{1}{c|}{2.16}   & 2.45   & \multicolumn{1}{c|}{2.27}   & \multicolumn{1}{c|}{3.20}   & \multicolumn{1}{c|}{3.23}   & 3.39   & \multicolumn{1}{c|}{1.89}  & 3.53   \\ \hline
			B2S\_max & \multicolumn{1}{c|}{62}    & \multicolumn{1}{c|}{76}     & \multicolumn{1}{c|}{89}     & 111    & \multicolumn{1}{c|}{63}    & \multicolumn{1}{c|}{70}     & \multicolumn{1}{c|}{145}    & 121    & \multicolumn{1}{c|}{2756}   & \multicolumn{1}{c|}{1727}   & \multicolumn{1}{c|}{1781}   & 2073   & \multicolumn{1}{c|}{36}    & 139    \\ \hline
			S2B\_avg & \multicolumn{1}{c|}{1.41}  & \multicolumn{1}{c|}{1.35}   & \multicolumn{1}{c|}{1.53}   & 1.86   & \multicolumn{1}{c|}{2.99}  & \multicolumn{1}{c|}{1.90}   & \multicolumn{1}{c|}{1.94}   & 2.17   & \multicolumn{1}{c|}{2.72}   & \multicolumn{1}{c|}{3.02}   & \multicolumn{1}{c|}{2.89}   & 2.80   & \multicolumn{1}{c|}{1.67}  & 2.58   \\ \hline
			S2B\_max & \multicolumn{1}{c|}{25}    & \multicolumn{1}{c|}{74}     & \multicolumn{1}{c|}{82}     & 76     & \multicolumn{1}{c|}{46}    & \multicolumn{1}{c|}{106}    & \multicolumn{1}{c|}{106}    & 139    & \multicolumn{1}{c|}{525}    & \multicolumn{1}{c|}{892}    & \multicolumn{1}{c|}{872}    & 900    & \multicolumn{1}{c|}{73}    & 229    \\ \hline
		\end{tabular}
	}
	\label{tab:inlining_four}
	\vspace{-8pt}
\end{table}

Table \ref{tab:inlining_four} shows the inlining statistics of these four datasets. 
Among the first three datasets, dataset II has the highest inlining frequency, and dataset III has the largest inlining degree. In particular, we notice that sometimes a BFI in dataset III can inline more than 2000 ISFs, and an ISF can be inlined into more than 800 BFIs. This is caused by some extremely-long call chains in the cryptography module of \textit{Openssl} \cite{openssl}, and we noticed that long call chains are also common in the dataset I and II. And in dataset IV, we observe higher inlining ratios in the released binary compared with dataset I-III which are compiled with default Ox optimizations, indicating large applications may use additional custom optimizations to further ensure the efficiency of the released executable.

\textbf{Answering RQ1.2:} Generally speaking, inlining widely exists in different kinds of projects. And in some large application binaries, inlining may happen more frequently.

\subsection{\textbf{RQ1.3:} What Parts of Inlining Are Cared for in Different Applications?} 
\label{sec:three_classes}

Different applications care about different parts of inlining due to their objectives. For example, code search aims to find target functions with similar functionality of the query function; thus, it needs to search BFs originated from the same OSF when function inlining happens. OSS reuse detection and vulnerability/patch detection aim to search functions containing the code of the reused or vulnerable functions; thus, they need to search BFs containing the same ISFs. Here, we focus on the inlining under OSS reuse detection and vulnerability/patch detection, and analyze the existence and proportion of target SFs in the query BF.

As shown in Figure \ref{fig:inlining_in_II}, we divided the relation between query BF and target SF into three classes: NSF, OSF, and ISF. For example, OSF means that the target SF is the OSF of the query BF, while ISF means that the target SF is an ISF of the query BF. Similar to Figure \ref{fig:inlining_in_II}, in Figure \ref{fig:inlining_in_III}, OSF means that query BF and target BF share the same OSF while ISF means the NSF of target BF (which is an NBF) is an ISF of the query BF.

For OSS reuse detection, we noticed that when optimization goes from O0 to O3, ISF is gradually counting for the largest part, and NSF decreases from more than 60\% to nearly 2\%, which indicates the majority reuse between source projects will be inlined when applying high optimization, making it a new challenge for existing works.

For vulnerability/patch detection, when applying high optimization, OSF becomes the majority where NBFs and BFIs need to be matched. Besides, 30\% of vulnerable functions are also inlined into BFIs, and finding these BFIs is also a task of existing works.

\textbf{Answering RQ1.3:} Code search works cares about the BFIs with the same OSF, while OSS reuse detection and vulnerability/patch detection also care about the BFIs with reused or vulnerable ISFs. The proportion of reused or vulnerable ISFs is not rare. Thus, they are also needed to be considered when conducting matching.

\subsection{Answer to RQ1}
In summary, function inlining widely exists in high optimizations and different projects. For released binaries, the inlining ratios range from 35\% to more than 70\%, and a BFI inlines 2-4 ISFs on average. 
Code search works should consider the unequal semantics between functions that inlining has brought. 
Besides, 80\% of reused SFs and 30\% of vulnerable SFs are inlined to BFIs when applying high optimization. OSS reuse detection and vulnerability/patch detection should also care about the hidden risky functions inlined into binary functions.


\section{Performance of existing works under inlining}
\label{sec:experiments}


This section will carry out several evaluations on the performance of existing works under inlining. Directly, we first split existing binary code similarity analysis works into binary2source matching and binary2binary matching works. Then we set different objectives for different applications. As we illustrated in Section \ref{sec:three_classes}, code search only needs to match BFs with the same functionalities, while OSS reuse detection and vulnerability/patch detection need to match every reused function. Thus, in our evaluation, apart from BFIs with the same OSF that code search works are required to accomplish, existing OSS reuse detection works and vulnerability/patch detection works are also evaluated on finding BFIs containing reused or vulnerable ISFs.

As inlining always comes with high optimizations, excluding the influence from other added optimization options is necessary. To evaluate the effect merely from inlining, we first select SFs that can produce BFs under all compilation settings. Then we split these SFs into two classes: OSFs whose BFs compiled in high optimization are BFIs and NSFs whose BFs are all NBFs. The only difference for BFs generated from these two classes is whether inlining has happened. Thus evaluations conducted on these two split datasets only reflect the influence of inlining.

For all works selected for evaluation, we use its default setting in its code or its paper and present its direct output.

\subsection{Evaluation of Binary2source Matching Works}
For binary2source matching works, we select CodeCMR\cite{CodeCMR} targeting two application areas: code search and OSS (open-source-software) reuse detection. To suit the applications, we use dataset I to evaluate the task in code search and dataset II for OSS reuse detection, and here are the results:

\subsubsection{\textbf{Binary2source code search} }

We select CodeCMR because it is the state-of-the-art function level binary2source code searching work. And its application binaryai\cite{binaryai} has been released for public usage. We analyze the impact of inlining on the performance of binaryai at a downstream task of code search: retrieving similar SFs for the query BF.

\begin{figure*}[h]
	\centering
	\vspace{-10pt}
	\subfigure[Cross-optimization evaluation]{
		\begin{minipage}[t]{0.58\linewidth}
			\centering
			\includegraphics[width=3.4in]{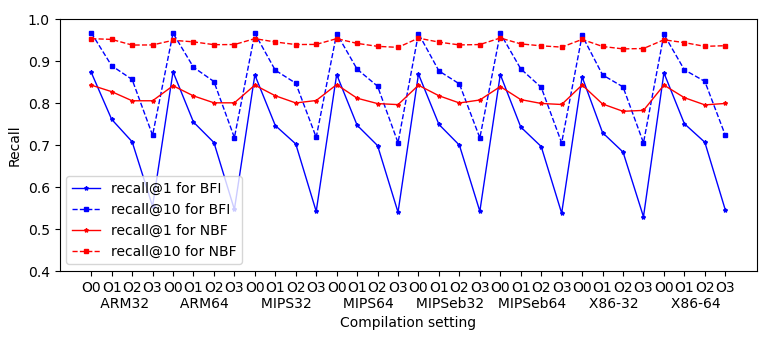}
			\label{fig:CodeCMR_evaluation}
			\vspace{-15pt}
		\end{minipage}%
	}%
	\centering
	\vspace{-10pt}
	\subfigure[Inlining-degree evaluation]{
		\begin{minipage}[t]{0.35\linewidth}
			\centering
			\includegraphics[width=2in]{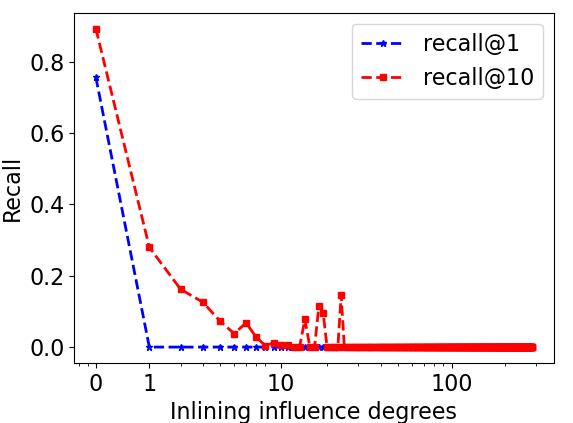}
			\label{fig:CodeCMR_inline_degree_to_recall}
			\vspace{-15pt}
		\end{minipage}%
	}%
	
	\caption{Evaluation of CodeCMR under inlining}
	\vspace{-12pt}
\end{figure*}

\textbf{Cross-optimization evaluation for CodeCMR.}  Figure \ref{fig:CodeCMR_evaluation} shows our evaluation of CodeCMR in BFIs and NBFs. We use the same metrics (recall@1 and recall@10) used in CodeCMR to evaluate its performance. As shown in Figure \ref{fig:CodeCMR_evaluation}, blue lines present the performance of matching OSFs with their BFIs, and red lines present the performance of matching NSFs with their NBFs. When matching functions without inlining, we can see a smooth curve for both recall@1 and recall@10, though they suffered a subtle loss when matching NSFs with NBFs from O0 to O3. But when matching functions with inlining, we can see a drastic loss from O0 to O3, about 30\% for recall@1 and 25\% for recall@10, which indicates that function inlining has brought a great challenge to CodeCMR.

\textbf{Inlining-degree evaluation for CodeCMR.} To further analyze the influence of inlining on CodeCMR, we rearrange the dataset into classes by their inlining influence degrees. In detail, we calculate the ratio between the total length of ISFs and the length of the OSF to define the influence degree that inlining has on matching a BFI with an OSF. And for matching pairs with different inlining degrees, we calculate their average recall@1 and recall@10 to form Figure \ref{fig:CodeCMR_inline_degree_to_recall}.

In Figure \ref{fig:CodeCMR_inline_degree_to_recall}, blue lines show the change of recall@1 and red lines for the recall@10. Inlining influence degree as ``1'' represents matching pairs that ISFs of BFI have a 1x to 2x length equal with the OSF of BFI. As shown in Figure \ref{fig:CodeCMR_inline_degree_to_recall}, both recall@1 and recall@10 decrease dramatically when inlining influence degree increase. In detail, recall@1 reaches nearly 80\% when the ISFs have a smaller length than the OSF but falls directly to 0 when the total length of ISFs is larger than the length of OSF. Recall@10 decreases when the inlining influence degree ranges from 0 to 10 and experiences a fluctuation between 10 and 30. But in general, when the inlining influence degree is larger than 1, it has become particularly difficult for CodeCMR to find the correct OSFs.

\subsubsection{\textbf{OSS reuse detection}} As we illustrated in Section \ref{sec:preliminary}, existing feature-based binary2source matching works cannot precisely detect software reuse at the function level. Thus, we extend CodeCMR to detect partial reuses.

\begin{minipage}[]{\textwidth}
	\vspace{7pt}
	\begin{minipage}[]{0.49\textwidth}
		\centering
		\makeatletter\def\@captype{table}\makeatother\caption{Function similarity for OSS reuse detection}
		\vspace{-5pt}
		\begin{tabular}{c|c|c|c|c}
			\hline
			similarity & O0                             & O1                            & O2                            & O3                            \\ \hline
			NSF        & 77.37\%                        & 73.31\%                       & 68.17\%                       & 58.12\%                       \\ \hline
			OSF        & 66.27\%                        & 71.42\%                       & 67.60\%                       & 58.08\%                       \\ \hline
			ISF        & {\color[HTML]{9A0000} 10.54\%} & {\color[HTML]{9A0000} 9.01\%} & {\color[HTML]{9A0000} 8.42\%} & {\color[HTML]{9A0000} 5.93\%} \\ \hline
		\end{tabular}
		\label{tab:CodeCMR_for_OSS}
	\end{minipage}
	\begin{minipage}[]{0.49\textwidth}
		\centering
		\makeatletter\def\@captype{table}\makeatother\caption{Function similarity for vulnerability detection}
		\vspace{-5pt}
		\begin{tabular}{c|c|c|c|c}
			\hline
			similarity & O0                             & O1                             & O2                             & O3                             \\ \hline
			NSF        & 89.36\%                        & 93.43\%                        & 92.68\%                        & 94.11\%                        \\ \hline
			OSF        & 82.69\%                        & 84.02\%                        & 85.92\%                        & 80.57\%                        \\ \hline
			ISF        & {\color[HTML]{9A0000} 46.16\%} & {\color[HTML]{9A0000} 63.39\%} & {\color[HTML]{9A0000} 64.34\%} & {\color[HTML]{9A0000} 55.80\%} \\ \hline
		\end{tabular}
		\label{tab:vulseeker_cross_class}
	\end{minipage}
	\vspace{10pt}
\end{minipage}

Table \ref{tab:CodeCMR_for_OSS} shows the results of CodeCMR in detecting three types of reuse corresponding to three classes in Section \ref{sec:three_classes}. When matching OSFs with BFIs, CodeCMR returns a similarity a little lower than matching NSFs with NBFs. However, when searching ISFs for BFIs, CodeCMR can only return a similarity of less than 10\% on average. As a result, CodeCMR can detect most NSFs and OSFs reused in NBFs and BFIs, but missed most ISFs that are inlined into BFIs.

\subsection{Evaluation of Binary2binary Matching Works}

For binary2binary matching works, we select SAFE \cite{safe} for code search, Vulseeker \cite{vulseeker} for vulnerability detection, and BinXray \cite{Binxray} for patch presence test. SAFE is evaluated on dataset I, while Vulseeker and BinXray are evaluated on dataset III.

\subsubsection{\textbf{Binary2binary code search}} 

We select SAFE for code search as it can represent the common workflow of most recent learning-based methods: training a model to produce embedding, input functions to obtain embeddings, and comparing embeddings for similarities.

\begin{figure}[htbp]
	\centering
	\vspace{-5pt}
	\begin{minipage}[t]{0.48\textwidth}
		\centering
		\includegraphics[width=0.9\textwidth]{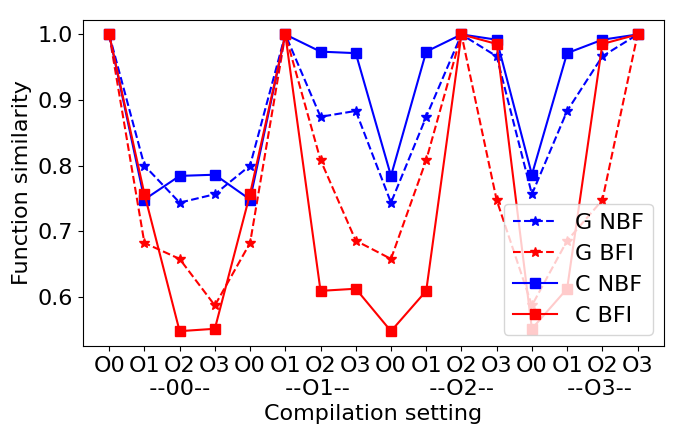}
		\vspace{-5pt}
		\caption{Cross-optimization function similarity for SAFE}
		\vspace{-10pt}
		\label{fig:safe_for_opt}
	\end{minipage}
	\begin{minipage}[t]{0.48\textwidth}
		\centering
		\includegraphics[width=0.85\textwidth]{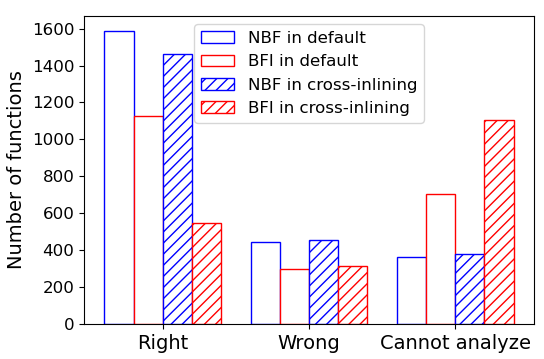}
		\vspace{-5pt}
		\caption{Cross-inlining results of BinXray}
		\label{fig:BinXray_cross_inlining}
	\end{minipage}
	\vspace{-10pt}
\end{figure}


Figure \ref{fig:safe_for_opt} shows the function similarity that SAFE obtained for cross-optimization code search. To reveal the impact of function inlining, we divide the matching cases into normal cases, which only compares two NBF and inlining cases where at least one function is a BFI. In Figure \ref{fig:safe_for_opt}, we use blue lines to represent the similarity in normal matching and red lines for inlining matching. Besides, we use `G' for GCC and `C' for Clang. And x-coordinate shows the detailed optimization setting of the experiments. For example, the first O1 with ``--O0--'' below represents searching query BFs generated by O0 in target BFs generated by O1.

In general, SAFE suffers a 20\% loss in similarity when detecting NBFs at the normal cross-optimization task (without inlining), but the loss increase to more than 40\% when detecting BFIs (with inlining). Interestingly, we noticed that inlining in clang has a greater impact on SAFE than gcc, and this difference becomes more obvious when crossing O0, O1 to O2, O3.

\subsubsection{\textbf{Vulnerability detection}} We select Vulseeker for vulnerability detection. To see the effect of inlining, we recompile dataset III with ``-fno-inline'' to create a dataset without inlining (dataset III-NI). Then we use vulnerable functions in dataset III-NI to generate fingerprints and functions in dataset III as the target. Results are shown in three types corresponding to three classes in Section \ref{sec:three_classes}.

As shown in Table \ref{tab:vulseeker_cross_class}, Vulseeker obtained a similarity of more than 90\% for normal matching cases, a similarity of 80\% when searching BFI, but a similarity of less than 65\% when searching inlined functions. As lower similarities indicate bigger difficulties in obtaining the target function, BFIs, which inline vulnerable ISFs, are very hard to find.

\subsubsection{\textbf{Patch presence test}} We select BinXray for patch presence test. As BinXray targets detecting patches at the same optimization, we conduct two experiments: default and cross-inlining. The default experiment generates fingerprints from binary functions compiled in O2 with inlining and search in binaries compiled in the same setting. The cross-inlining experiment generates fingerprints from binary functions compiled in O2 without inlining but searches binary functions compiled in O2 with inlining. Here are the results.

Figure \ref{fig:BinXray_cross_inlining} shows the results of default experiment and cross-inlining experiment. The number of rightly identified functions drops when detecting NBFs in the cross-inlining experiment compared with the default experiment, and this gap grows much larger for BFIs. Only about 2/5 of the vulnerable BFIs are rightly classified in the cross-inlining experiment, while 3/5 fall to ``cannot analyze''.

Interestingly, although the rightly identified cases decrease dramatically, wrongly identified only increase a little. Instead, more cases become hard to be analyzed. That is because the analysis of BinXray relies on its successful identification of the patch boundary. For BFIs with vulnerable OSF that we mention in section \ref{sec:three_classes}, BinXray cannot find the similar patch boundary derived from dataset III-NI. 

Besides, another problem is that BinXray regards the functions having the same name with vulnerable function as the suspicious function to conduct detecting. Thus BFIs which has inlined vulnerable ISFs will all be missed. Note that these BFIs with vulnerable ISFs account for 30\% when compiling in O3 as shown in Section \ref{sec:three_classes}.

\subsection{Answering to RQ2}
In summary, code search works suffer a 20\%-30\% performance loss when detecting functions with inlining. Ignorance of semantics from inlined functions reveals their inherent defect in detecting function similarities. Besides, OSS reuse detection and vulnerability/patch detection works cannot detect inlined functions. Thus, attackers or plagiarists can inline these risky functions into binary functions to bypass the detection.

\section{EVALUATION of existing inlining-simulation strategies}
\label{sec:inline_pattern}

In section \ref{sec:experiments}, we have shown that existing works suffer from the challenges that inlining brings. To resolve these challenges, existing works propose inlining-simulation strategies to simulate the inlining results of compilers. In this section, we will introduce existing inlining-simulation strategies and evaluate their effectiveness against function inlining in two fundamental tasks.

\subsection{Introduction of Existing Inlining-simulation Strategies}

We first introduce the strategies used in two existing works: Bingo\cite{bingo} and  ASM2Vec\cite{asm2vec}.

\textbf{Bingo.} Bingo proposes a selective inlining-simulation strategy to recursively expand callee functions for more precise similarity detection. To decide whether a callee should be inlined, Bingo listed 6 patterns for inlining conduction, and we summarize them into the following 5 cases where 2-5 are towards UD (User-Defined) functions.

1. if the callee function is a library function, inline it.

2. If the UD callee function and its caller recursively call each other, inline it.

3. If the UD callee function only calls library functions, and more than half of library functions are not termination functions (e.g., exit and abort), inline it.

4. If the UD callee function only calls library functions, and more than half of library functions are termination functions, do not inline it.

5. If the UD callee function calls other UD functions, the inlining is decided by the following equation \ref{equ:bingo}.

\vspace{-10pt}
\begin{equation}
	\alpha(f_c) = outdegree(f_c)/(outdegree(f_c)+indegree(f_c))\label{equ:bingo}
\end{equation}

In equation \ref{equ:bingo}, in-degree and out-degree refer to the call relation between the callee $f_c$ and other UD functions in the FCG (Function call graph). Bingo considers that the lower the value of  $\alpha(f_c)$, the more likely the callee should be inlined. Thus when $\alpha(f_c)$ is lower than a threshold, this callee will be inlined. In Bingo's original setting, this threshold is set to 0.01. 

\textbf{ASM2Vec.} The inlining-simulation strategy of ASM2Vec employs the strategy of Bingo, but with some modifications and additional filters. On the one hand, ASM2Vec change the recursively inlining-simulation strategy into a one-layer inlining-simulation strategy, where they only expand the first-order callees. On the other hand, they defined a metric to remove lengthy callees, as shown in equation \ref{equ:asm2vec}.

\vspace{-10pt}
\begin{equation}
	\delta(f_s, f_c) = length(f_c)/length(f_s) \label{equ:asm2vec}
\end{equation}

In equation \ref{equ:asm2vec}, $f_c$ represents for the callee function and $f_s$ for the caller. The length of the function is measured by the number of lines of instructions. When $\delta$ is less than 0.6 or $f_s$ is comprised of less than 10 lines of instructions (considered as a wrapper function), inlining will happen.

\subsection{Evaluating Strategies on Finding Similar BFIs}
\label{sec:inlining_strategy_in_b2b}

As finding BFIs with the same OSF is the main objective of existing binary code similarity analysis works, we first evaluate their effectiveness in finding similar BFIs. According to their settings, the inlining-simulation strategies will be conducted for both the query BF and target BF, and the $BFI'$s produced by the inlining-simulation strategies help find similar BFIs.

To evaluate their effectiveness, we implement the inlining-simulation strategies of Bingo and ASM2Vec and apply them to dataset I. For each BF, Bingo and ASM2Vec will generate the $BFI'$ using their inlining-simulation strategies. In detail, we use IDA\cite{IDAPro} to disassemble binaries and extract features used for inlining selection. And we provide two metrics to evaluate the effectiveness of their strategies.

\vspace{-10pt}
\begin{equation}
	cost =BFI_1' - BF_1 + BFI_2' - BF_2 \label{equ:metric_cost}
\end{equation}

\vspace{-10pt}
\begin{equation}
	similarity = \frac{BFI_1' \cap BFI_2'}{BFI_1' \cup BFI_2'}  \label{equ:metric_sim}
\end{equation}

In Equation \ref{equ:metric_cost} and \ref{equ:metric_sim}, $BF_1$ and $BF_2$ are two BFs that are need to be compared, and $BFI_1'$ and $BFI_2'$ are the $BFI'$s generated by the inlining-simulation strategies proposed by Bingo and ASM2Vec.  Equation \ref{equ:metric_cost} calculates the number of inlined BFs for both $BF_1$ and $BF_2$, and Equation \ref{equ:metric_sim} calculates the similarity of produced $BFI_1'$ and $BFI_2'$. And we measure the $BFI'$ in equation \ref{equ:metric_cost} by the BFs it contains and $BFI'$ in equation \ref{equ:metric_sim} by the SFs it contains.

\begin{figure}[htbp]
	\centering
	\vspace{-5pt}
	\includegraphics[width=0.65\textwidth]{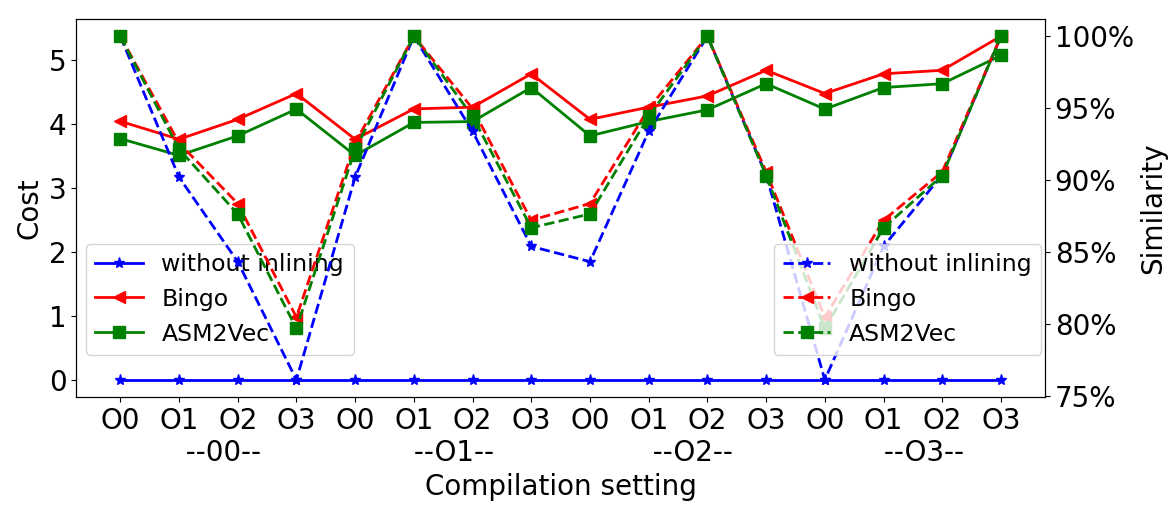}
	\vspace{-10pt}
	\caption{Effectiveness of existing inlining-simulation strategies on binary2binary matching}
	\label{fig:b2b_evaluation}
	\vspace{-12pt}
\end{figure}

Figure \ref{fig:b2b_evaluation} shows the evaluation results of Bingo and ASM2Vec. To facilitate comparison, we also present the cost and similarity without applying inlining strategies (without inlining in Figure \ref{fig:b2b_evaluation}). The dotted line represents the similarity and the solid line for the cost. In general, compared with not applying strategies, Bingo averagely selects 4-5 BFs for inlining to obtain a 5\% increase in its obtained similarity. And ASM2Vec can reduce 0.25 BFs for inlining compared to Bingo while losing 1-2\% in the similarity.

From the results, we can notice that Bingo and ASM2Vec are selective towards callees as recursively inlining all callees will cause a cost to inline nearly 55 callees while they only need to inline 4-5 callees. And their effectiveness is quite visible --- a 5 \% increase in similarity.
However, there is still a large space to improve, as when detecting functions cross O0 to O3, existing inlining-simulation strategies can only help guarantee that 80\% of SFs in $BFI_1'$ and $BFI_2'$ are matched. And we noticed that some BFs, which both are the callees of $BF_1$ and $BF_2$, are also inlined by $BFI_1'$ and $BFI_2'$, helping increase the function similarity.

In general, Bingo and ASM2Vec can help existing works improve their accuracy in finding similar BFIs, but there is still a large space for improvement.

\subsection{Evaluating Strategies on Finding Inlined Functions} 

In our evaluation of existing binary similarity analysis works, we noticed new requirements in some specific areas, including OSS reuse detection, vulnerability detection, and patch presence test, where inlined reused and inlined vulnerable functions also need to be detected. In this section, we will evaluate the ability of Bingo and ASM2Vec in finding inlined functions.

\textbf{Bingo.} To facilitate evaluation, we first apply Bingo to dataset I and analyze its inlining decisions towards BFs. As Bingo has 3 cases (2, 3, 5) that callees (UD functions) will be inlined, to evaluate the effectiveness of its inlining-simulation strategies, we summarize the inlining decisions under the above 3 cases in O0-O3, which is shown in Table \ref{tab:bingo_b2s}.

\begin{table}[htbp]
	\caption{Inlining decision of Bingo for BFs}
	\centering
	\vspace{-7pt}
	\begin{tabular}{cc|c|c|c|c|c|c}
		\hline
		\multicolumn{1}{c|}{Kind}                 & Case     & O0-O1                        & O0-O2                        & O0-O3                        & O1-O2                        & O1-O3                        & O2-O3                       \\ \hline
		\multicolumn{1}{c|}{\multirow{3}{*}{BFNN(BF Not Needed)}} & $\alpha$ & 20.56\%                      & 21.57\%                      & 21.79\%                      & 20.26\%                      & 20.14\%                      & 25.07\%                     \\ \cline{2-8} 
		\multicolumn{1}{c|}{}                     & L        & 68.56\%                      & 45.48\%                      & 42.53\%                      & 50.53\%                      & 46.68\%                      & 57.47\%                     \\ \cline{2-8} 
		\multicolumn{1}{c|}{}                     & R        & 4.22\%                       & 4.57\%                       & 4.59\%                       & 5.56\%                       & 5.50\%                       & 3.21\%                      \\ \hline
		\multicolumn{1}{c|}{\multirow{3}{*}{BFN(BF Needed)}} & $\alpha$ & 0.14\%                       & 0.32\%                       & 0.32\%                       & 0.15\%                       & 0.15\%                       & 0.00\%                      \\ \cline{2-8} 
		\multicolumn{1}{c|}{}                     & L        & 6.41\%                       & 27.85\%                      & 30.49\%                      & 22.83\%                      & 26.71\%                      & 8.84\%                      \\ \cline{2-8} 
		\multicolumn{1}{c|}{}                     & R        & 0.12\%                       & 0.22\%                       & 0.28\%                       & 0.69\%                       & 0.81\%                       & 5.42\%                      \\ \hline \hline
		\multicolumn{2}{c|}{coverage}                        & \multicolumn{1}{l|}{78.51\%} & \multicolumn{1}{l|}{67.65\%} & \multicolumn{1}{l|}{60.68\%} & \multicolumn{1}{l|}{72.42\%} & \multicolumn{1}{l|}{65.01\%} & \multicolumn{1}{l}{74.33\%} \\ \hline
	\end{tabular}
	\label{tab:bingo_b2s}
	\vspace{-5pt}
\end{table}

In table \ref{tab:bingo_b2s}, R, L, and $\alpha$ respectively represents inlining for \textbf{R}ecursive call in case 2, inlining for callees with only \textbf{L}ibrary call in case 3 and inlining determined by $\boldsymbol{\alpha}$ in case 5. BFNN represents BFs that do not need to be inlined, while BFN represents BFs needed. To facilitate analysis, we only present results of comparing BFs generated by low optimization to BFs generated by high optimizations, and we analyze its inlining conduction for the BFs in low optimizations. For example, the first value $20.56\%$ indicates that, when matching BFs generated by O0 to BFs generated by O1, among all inlined functions, $20.56\%$ are BFs which are not needed to be inlined but inlined, as it falls to case 5 and its $\alpha$ satisfies Bingo's requirements.

For an effective inlining-simulation strategy, the percentage of BFNN should be low as they do not need to be inlined, and the percentage of BFN should be high as they need to be inlined. But from table \ref{tab:bingo_b2s}, we can see that nearly 3/4 of inlined functions are BFNNs. To be more specific, among the three metrics, we noticed that only L had selected a large amount of BFNs while $\alpha$ and R rarely select BFNs. Bingo's strategies waste a lot of cost in inlining unnecessary callees.

Besides, table \ref{tab:bingo_b2s} also shows the coverage of needed inlined functions in each matching. When the compilation setting is not so diverse, Bingo can cover more than 70\% inlined functions. But when comparing O0 to O3, 40\% of inlined functions remain uncovered, making the results less complete.

\textbf{ASM2Vec.} ASM2Vec proposed two measures to reduce the functions Bingo chooses to inline. They have a effect in reducing the inlining candidates (see Section \ref{sec:inlining_strategy_in_b2b}). However, we argue that their restrictions are still so strict that they would suffer additional misses of ISFs. To provide a comprehensive understanding of boundaries for inlining, we conduct a study to analyze inlining from SFs that comprise the final BFIs.

\begin{figure*}[htbp]
	\centering
	\vspace{-15pt}
	\subfigure[Depths of BFIs]{
		\begin{minipage}[t]{0.33\linewidth}
			\centering
			\includegraphics[width=1.9in]{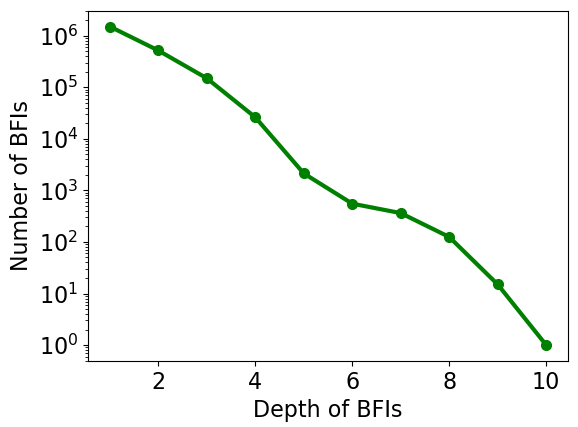}
			\label{fig:asm2vec_inlining_depth}
			\vspace{-19pt}
		\end{minipage}%
	}%
	\centering
	\subfigure[Lengths of ISFs]{
		\begin{minipage}[t]{0.33\linewidth}
			\centering
			\includegraphics[width=1.9in]{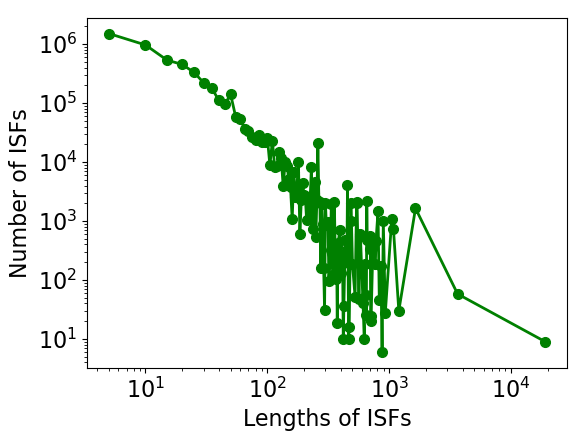}
			\label{fig:b2s_inlinined_lengths}
			\vspace{-19pt}
		\end{minipage}%
	}%
	\subfigure[Ratio of ISF/OSF]{
		\begin{minipage}[t]{0.33\linewidth}
			\centering
			\includegraphics[width=1.9in]{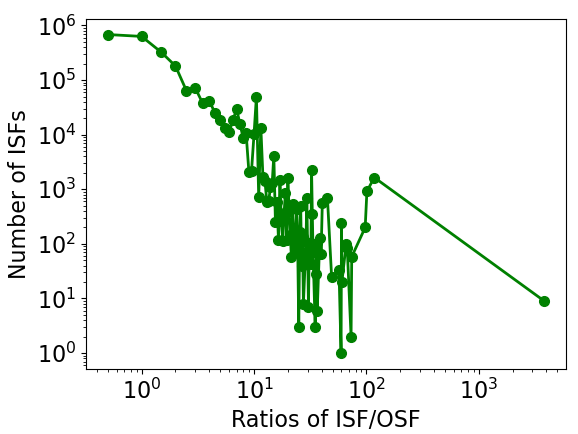}
			\label{fig:asm2vec_len_ratio}
			\vspace{-19pt}
		\end{minipage}%
	}%
	\vspace{-10pt}
	\caption{Distribution of metrics for inlined source functions}
	\vspace{-15pt}
\end{figure*}

Figure \ref{fig:asm2vec_inlining_depth} shows the distribution of depth for BFIs. We take the example in Figure \ref{fig:linux_inline_example_all} (d) to illustrate the meaning of depth. In Figure \ref{fig:linux_inline_example_all} (d), BFI \textit{netlink\_sendmsg} has inlined four source functions, including \textit{scm\_set\_cred} and \textit{scm\_send}. And \textit{scm\_send} is called by OSF \textit{netlink\_sendmsg} and \textit{scm\_set\_cred} is called by ISF \textit{scm\_send}, where we recognize the depth of \textit{scm\_send} as 1 and the depth of \textit{scm\_set\_cred} as 2. Finally, we use the maximum of depths (i.e.,2) as the depth of this BFI.
As shown in Figure \ref{fig:asm2vec_inlining_depth}, hundreds of thousands of BFIs inline ISFs with a calling depth of 2-3, and hundreds of BFIs inline ISFs even through 8 layers of the call graph. 30\% of the BFIs have a depth of more than 1 layer.

Figure \ref{fig:b2s_inlinined_lengths} shows the distributions of ISFs lengths. Though lots of ISFs have fewer than 10 lines, we noticed that tens of thousands of ISFs are above 100 lines. Even thousands of ISFs with a length above 1000 are still inlined. Interestingly, the largest ISF is ``\textit{aarch64\_opcode\_lookup\_1}'' with 19083 lines of code which are inlined by the OSF ``\textit{aarch64\_opcode\_lookup}'' with less than 10 lines of code. And this inlining happens during the compilation of Binutils compiling with GCC in O1. This is actually caused by the ``-finline-functions-called-once'' option as the ISF ``\textit{aarch64\_opcode\_lookup\_1}'' is only called by the OSF ``\textit{aarch64\_opcode\_lookup}''.

Figure \ref{fig:asm2vec_len_ratio} shows the distribution of ratios between ISF and OSF. For all ISFs, we noticed that ratios of ISF/OSF range from 0 to more than 1000. Only less than 1/3 of ISFs have a less than 0.6 of $\delta$ that will be inlined according to the ASM2Vec strategies, while 50\% of the ISFs have a larger length than their OSF. 

When using one-layer inlining and removing candidates whose $\delta >=0.6$, ASM2Vec additionally misses ISFs. The missing of these ISFs not only influences the performance in finding the OSF of BFI but also its ability to detect BFIs that have inlined reused or vulnerable ISFs.

\subsection{Answer to RQ3}

Bingo and ASM2Vec can help find similar BFIs, but there is still a large space for promotion. When applying to detecting inlined functions, they miss nearly 40\% of the inlined functions. More effective inlining-simulation strategies are needed to resolve the effect of function inlining on binary code analysis works.

\section{Towards more effective strategies}
\label{sec:suggestion}

This section will discuss how to design a more effective inlining-simulation strategy to help further reduce the inlining cost and find more ISFs. In detail, we first recover the matching patterns that inlining brings, and based on the findings, we propose several suitable measures to help develop a more effective strategy.

\subsection{\textbf{Investigation of Matching Patterns Under Inlining}}

\begin{figure}[htbp]
	\centering
	\vspace{-5pt}
	\includegraphics[width=0.8\textwidth]{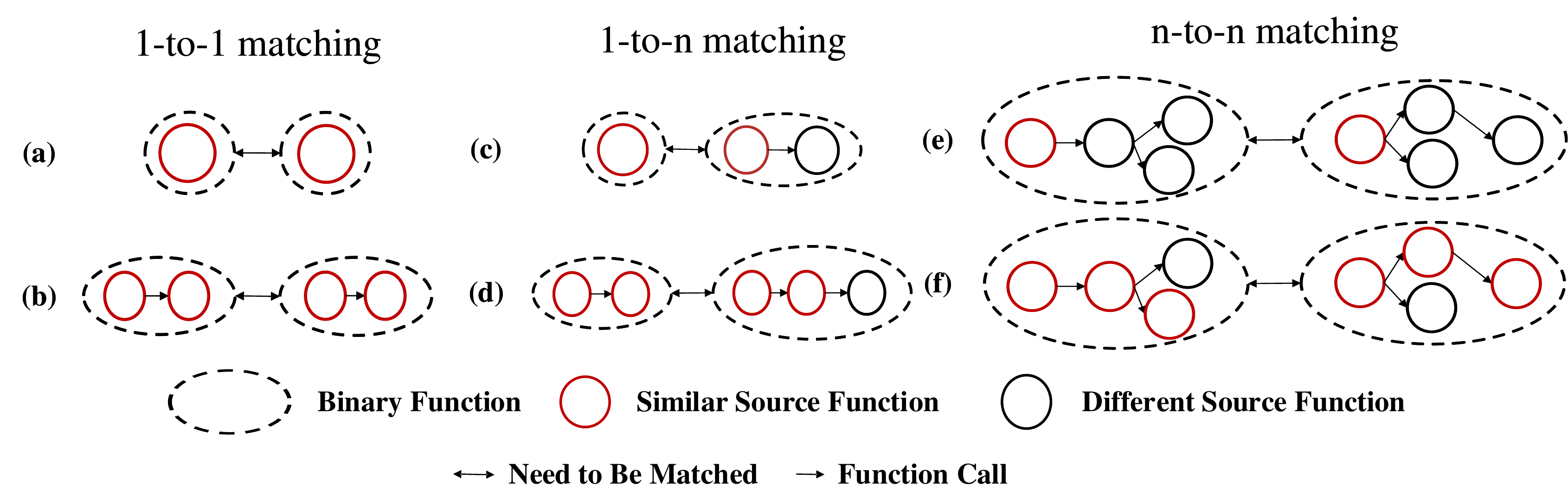}
	\vspace{-5pt}
	\caption{Binary2binary matching patterns under inlining}
	\label{fig:b2b_matching_patterns}
	\vspace{-5pt}
\end{figure}

The matching pattern inlining brings to binary2source matching is ``1-to-n'' compared with the existing ``1-to-1'' matching mechanism. However, this becomes much more complex for binary2binary matching. Here, we summarize the patterns of function-level matching under function inlining, as shown in Figure \ref{fig:b2b_matching_patterns}. 


\textbf{1-to-1 matching.} ``1-to-1'' matching is usually the pattern that most existing binary2binary work applies. Figure \ref{fig:b2b_matching_patterns} (a) just shows the matching without function inlining. And another matching with function inlining is also ``1-to-1'' matching, where two BFs have inlined the same ISFs and thus have equal semantics.

\textbf{1-to-n matching.} ``1-to-n'' matching is similar to the pattern in binary2source matching. Figure \ref{fig:b2b_matching_patterns} (c) shows the similar matching with binary2source matching. Apart from that, as shown in Figure \ref{fig:b2b_matching_patterns} (d), matching the BFI \textit{A} with the BFI \textit{B}, which has inlines all ISFs in \textit{A} and other ISFs, is also an ``1-to-n'' matching. 

\textbf{n-to-n matching.} ``n-to-n'' matching reveals the most complex matching patterns when conducting binary2binary matching. Figure \ref{fig:b2b_matching_patterns} (e) and (f) shows two classes of ``n-to-n'' matching. In detail, (e) shows two BFIs which have inlined different ISFs and the common semantics of these two BFIs only come from their shared OSF. (f) describe the BFI pairs which share the OSF and other ISFs.


Existing inlining-simulation strategies are conducted both at the query and target functions, so they can resolve these three matching patterns. However, we recovered that not all these patterns are common in all matching cases. Thus, we could focus only on one or two particular patterns to resolve the challenges that inlining brings.

\begin{table}[h]
	\vspace{-5pt}
	\caption{Distribution of binary2binary matching patterns}
	\centering
	\vspace{-7pt}
	\renewcommand{\arraystretch}{1.3}
	\begin{tabular}{c|c|c|c|c|c|c}
		\hline
		Pattern & a       & b      & c       & d      & e      & f      \\ \hline
		Percent          & 63.80\% & 6.33\% & 25.70\% & 3.11\% & 0.82\% & 0.21\% \\ \hline
	\end{tabular}
	\vspace{-5pt}
	\label{tab:b2b_distribution}
\end{table}

Table \ref{tab:b2b_distribution} shows the proportion of binary2binary matching patterns by comparing BFs originated from the same OSF but compiled with different architectures, compilers, and optimizations in dataset I. Interestingly, pattern e and f account for only 1\% even with distant compilation settings. Instead, ``1-to-1'' matching accounts for the most common part as 70\%, and ``1-to-n'' matching accounts for the rest 29\%. To be more specific, comparing BFIs compiled in low optimizations with BFIs compiled in high optimizations are mostly cases falling to the pattern (c) and (d), which indicates the inlining is usually cumulative when optimization increases.

\subsection{\textbf{Suggestions for Designing More Effective Strategies}} 
Considering the evaluation results of Bingo and ASM2Vec and proportion of matching patterns in binary2binary matching, we find that existing strategies have three main shortcomings: 

First, existing strategies conduct inlining for all cases on two sides. However,  70\% of matching cases do not need inlining, while 29\% only need inlining at one side. Only 1\% of matching cases need inlining at two sides. 
Second, inlining unnecessary functions will bring difficulties in identifying the true match. For example, Bingo inlines some functions both in two functions, where different BFs with the same callees after inlining will also become similar. 
Third, many ISFs are missed as the side-effect of reducing the inlining candidates. For example, the restriction of Bingo makes it miss nearly 40\% of inlined functions, and ASM2Vec applies two restrictions for candidates resulting in more missed ISFs. As a result, it cannot help detect BFIs with vulnerable ISFs inlined.
To overcome the shortcomings listed above, we propose several suggestions to reduce the inlining cost and help find the inlined functions.

\textbf{Necessary preprocessing to reduce candidates. } In binary2source matching, user-forced inlining candidates should be first identified. For example, functions with ``$\_\_attribute\_\_(always\_inline)$''  will be inlined. Functions with ``$\_\_attribute\_\_(noinline)$'' will not be inlined. Thus, these functions can be determined whether to inline before applying strategies. Further, callees of not inlined functions can be filtered, and callees of inlined ones can be added to candidates.

\textbf{Inspection of compilation settings to decide whether to apply strategies.} Leveraging some works designed for toolchain provenance recovery \cite{rosenblum2011recovering, rahimian2015bincomp, bardin2021compiler, otsubo2020glassesx, tian2021fine}, the compilation setting of binaries can be inferred. Then we could decide whether to apply strategies according their compilation settings. For example, in binary2source matching, strategies are not needed if the binary is generated without inlining (such as ``-O0''). In binary2binary matching, strategies are not needed if binaries are generated in the same compilation setting. 

\textbf{Relaxing strategies of Bingo and ASM2Vec to recover more ISFs.} Section \ref{sec:inline_pattern} reveals the effectiveness of Bingo and ASM2Vec in reducing candidates but at the cost of missing inlined functions. We find that the strategy of Bingo and ASM2Vec is too strict. We can relax their restriction to obtain more inlined functions. For example, changing the restriction from $\alpha < 0.01$ to $\alpha < 0.5$ helps increase the similarity by 10\% while only inlining 5 more BFs and increasing the depth of BFIs from 1 to 2 can improve the coverage of ISFs from 70\% to nearly 93\%.

\textbf{Incremental inlining-simulation strategy to help locate ISFs.} Considering that the most matching pattern is ``1-to-n'' when inlining happens, we do not need to expend all suitable callees at both sides. Interestingly, we find that we can conduct an inlining trail to filter inlining candidates if the method can distinguish between more similar and less similar function pairs. For function pairs such as the example in Figure \ref{fig:b2b_matching_patterns} (c), we named that NBF as A and BFI as B. Initially, we discover many inlining candidates for both A and B. For A, we take a trail to inline one of the candidates into A to form A with inlining. Then we compare the similarity between A and B to the similarity between A with inlining and B. If the similarity increase with inlining, we regard it as a suitable candidate and continue. Otherwise, the candidate, along with its callees, can be filtered. By conducting the inlining trail incrementally, we can quickly reduce the unnecessary branches and focus on some branches with deeper depth.


\section{Threats to validity}
\label{sec:disscusion}


\textbf{Dataset labeling method.} As our dataset labeling method is based on existing tools, the accuracy of dataset labeling depends on the accuracy of existing tools. In fact, we notice that there are difficulties for \textit{understand} in parsing source entities, so we use a conservative approach to label the source code. In detail, we run \textit{understand} in its ``strict'' mode where it will accurately identify the entity or add it to the unresolved list. And we manually label the unresolved entities to complement them.

\textbf{Inspecting inlining from the generated binaries.} Some readers may wonder why we inspect inlining from the generated binaries instead of the compiler designs. There are three reasons. First, understanding the inlining strategies from compilers' source code seems feasible, but this requires great manual effort. Moreover, as the inlining strategies of different compilers in different versions will differ, we regard directly extracting inlining statistics from binary as a more general way. Second, inlining results rely on not only the inlining strategies but also the features of the source project. For example, the inlining decision of GCC depends on the total cost of all inlined call-sites; thus, without the exact project, the inlining decision will not be clear. Third, we also combine some inlining strategies with the inlining statistics in binary. We think using strategies to assist in understanding function inlining is a good complement for inspecting function inlining in real binaries.


\textbf{Metrics for evaluating existing works.} In Section \ref{sec:experiments}, we follow the following principles to determine the metrics for evaluating existing works. If the work has used a metric to evaluate its effectiveness, we directly use it. Otherwise, we use its direct output, such as the similarities between functions, to present its results. We do not choose to implement a new metric because the function similarities are enough to present their effectiveness under inlining, as the metrics used to evaluate downstream tasks are all based on function similarity.

\vspace{-2pt}
\section{Conclusion}

For the first time, our work investigates the effect of function inlining on binary similarity analysis research. 
Four datasets are constructed, and an automatic identification method is proposed to facilitate the analysis of function inlining. Our analysis finds that 36\%-70\% binary functions have functions inlined in high optimizations, while most binary code similarity still regards function mapping as ``1-to-1''. This mismatch causes 30\% loss in performance during code search and 40\% loss during vulnerability detection. Moreover, inlined functions are nearly all ignored during OSS reuse detection and patch presence test. Furthermore, existing inlining-simulation strategies usually produce less complete inlining results, leaving the binary similarity analysis under function inlining still challenged.

We propose several suggestions to design a more effective inlining-simulation strategy, but we still regard it as unfinished work. We hope that more researchers can pay attention to function inlining and conduct more research to resolve function-inlining-related issues.

\bibliographystyle{ACM-Reference-Format}
\bibliography{references}

\end{document}